# Metal-Silicene Interaction Studied by Scanning Tunneling Microscopy

Zhi Li[1], Haifeng Feng[1], Jincheng Zhuang[1], Na Pu[2], Li Wang[1], Xun Xu[1], Weichang Hao[2], Yi Du[1,*]

[1]Institute for Superconducting and Electronic Materials (ISEM), Australian Institute for Innovative Materials (AIIM), University of Wollongong, Wollongong, NSW 2500, Australia.

[2]Beihang University, 37 Xueyuan Rd, Haidian District, Beijing 10091, P. R. China.

* To whom correspondence should be addressed. Email: yi_du@uow.edu.au

**Abstract**

Ag atoms have been deposited on 3×3 silicene and √3×√3 silicene films by molecular beam epitaxy method in ultrahigh vacuum. Using scanning tunnelling microscopy and Raman spectroscopy, we found that Ag atoms do not form chemical bonds with both 3×3 silicene and √3×√3 silicene films, which is due to chemically inert surface of silicene. On 3×3 silicene films, Ag atoms mostly form into stable flat top Ag islands. In contrast, Ag atoms form nanoclusters and glide on silicene films, suggesting more inert nature. Raman spectroscopy suggests that there is more $sp^2$ hybridization in √3×√3 than in √7×√7/3×3 silicene films.

**1. Introduction**

Silicene, a two-dimensional (2D) monolayer honeycomb structure of silicon [1,2], has attracted intensive research interest due to its novel physical properties, including 2D topological insulator characteristics and quantum spin Hall effect (QSHE) [3-5]. Scanning tunneling microscopy (STM) and angle-resolved photoemission spectroscopy (ARPES) studies have verified the existence of a linear dispersion band structure near Fermi level in epitaxial silicene grown on Ag(111), which demonstrates this silicon-based 2D material is a new Dirac fermion system [6-8]. In contrast to the other 2D materials such as graphene, epitaxial silicene shows rich phases including √7×√7, 3×3 and √3×√3 (in respect to 1×1 silicene) etc. [2, 6-11], which is believed to be determined by the interaction with the substrate. As a result, various low-buckled configurations are formed in silicene corresponding to different phases. Consequently, the hybridization states in silicene films are expected to be a mixture of $sp^2$ and $sp^3$ hybridizations [12]. The buckled structure is dominated by the $sp^2$ to $sp^3$ ratio. Generally, it is believed √3×√3 silicene possesses more $sp^2$ state than √7×√7 and 3×3 silicene. This hypothesis helps to understand why physical and chemical characteristics of different silicene phases show great differences [12]. For example, √7×√7 and 3×3 silicene exhibit a highly chemical reactivity to oxygen, while √3×√3 silicene is relative chemical inert and remains its structures more than 24 hours in air [10-12]. In terms of electronic structures, Dirac cone either vanishes in epitaxial √7×√7 phase or is strongly twisted in 3×3 phase [8] due to strong interface hybridizations between Si in silicene and Ag in the substrate. In contrast, epitaxial √3×√3 silicene possesses the Dirac fermion characteristics with a lightly n-type doping from Ag(111) substrate [7]. Its surface properties has been theoretically studied [16], and very recently, been verified experimentally by a hydrogenation of 3×3 silicene[17]. Although these findings suggest that the hybridization between $s$-orbital and $p$-orbitals in silicene film determines their surface properties, the detailed study are still critically unclear and have not been investigated experimentally.

In this work, we have investigated the effect of silicene phases on the structure and mobility of Ag nanoclusters deposited on silicene films by using low-temperature STM. It was found that both 3×3 and √3×√3 silicene are free of new reconstructions, indicating the inert nature of these phases. However, the structures of Ag nanoclusters and islands are determined by silicene phases, which reflect distinct surface buckling configurations of 3×3 and √3×√3 silicene. The formation of Ag islands on 3×3 silicene indicates relatively strong interaction between Ag and 3×3 silicene, which is most likely induced by the high $sp^3$ component. The high mobility of Ag clusters on √3×√3 silicene indicates that the surface of √3×√3 silicene is more inert which is possibly because its hybridization state is mainly a $sp^2$ type. By using an *in-situ* Raman spectroscopy, we have further proved that the observed differences in surface properties of silicene phases are attributed to ratio of Si $sp^2$/ $sp^3$ hybridization configurations.

## 2. Experimental

This work is carried out on an ultrahigh vacuum (UHV) low temperature STM system, equipped with a molecular beam epitaxy (MBE) chamber and an *in-situ* Raman spectroscopy. High-quality silicene films growth and deposition of silver are conducted in MBE chamber before being transferred into STM for *in-situ* measurements. The Ag(111) substrate was cleaned by cycles of $Ar^+$ bombardment and annealing, then Si was evaporated onto Ag(111) substrate from a high-purity silicon wafer to form silicene films. The detailed growth method of silicene films is described elsewhere [13,14]. Ag was deposited from a standard Knudsen cell onto the as-grown silicene films at 150 K. All STM/STS and Raman spectroscopy measurements were performed at 77 K. The differential conductance d$I$/d$V$ spectra were acquired by using a standard lock-in technique with a 20 mV modulation at 937 Hz. The *in-situ* Raman measurements were carried out on the same silicene samples in UHV at 77 K.

## 3. Results and discussion

Figure 1(*a*) shows STM image of three phases of silicene. The top part is 3×3 silicene which dominates the first layer silicene growth directly on Ag(111) substrate. Due to similar formation energy, √7×√7 silicene usually appears together with 3×3 silicene but with significantly lower coverage. After Ag(111) substrate been fully covered by 3×3 and √7×√7 silicene, √3×√3 silicene starts to grow onto these two phases. Thus, by controlling coverage of silicene, the uniform film of 3×3 silicene (with small percentage of √7×√7 silicene) or √3×√3 silicene can be precisely prepared. The typical high-resolution STM images of these three silicene phases are present in Figure 1(*b*)-1(*d*). Due buckling differences [9], all three phases are distinctive from each other in STM image. Although 1×1 silicene possesses a honeycomb structure, the strong buckling and interaction with Ag(111) substrate alter the configuration of silicon atom arrangement, inducing a complex structure of 3×3 silicene (Figure 1(*b*)) rather than honeycomb structure. However, in √3×√3 silicene (Figure 1(*d*)), honeycomb structure is preserved, suggesting the dominating hybridization is possibly

$sp^2$, similar to graphene. In contrast to the intact and well-ordered structure of 3×3 and √3×√3 phases, √7×√7 phase always shows many structural defects (Figure 1(*a*) and (*c*)), indicating a poorer structure stability in consistent with the minor coverage of this phase. In this report, we mainly focus on the dominating phases of 3×3 silicene and √3×√3 silicene to figure out the different hybridization configurations between them.

Figure 2(*a*) shows STM image of 3×3 silicene film after deposition of Ag. Most Ag atoms form flat top islands. As shown in the inset figure, the height profile taken along the solid line shows the island has a uniform height about 6 Å. Beside of these islands, there are few Ag clusters, which are probably due to the low deposition temperature (150 K). But no new surface reconstruction has been observed after Ag deposition, indicating a relatively inert surface comparing with 7×7 reconstruction of Si(111) surface [18]. This suggests hybridization state of 3×3 silicene is different from pure $sp^3$ type. High-resolution image (Figure 2(*b*)) shows structure of Ag island has a same periodicity with 3×3 silicene. Similar reconstruction in Ag is not reported on Ag single crystal as far as we know. Spectra (Figure 2(*c*)) show the prominent peak at -0.5 V in 3×3 silicene, which is probably due to the saddle point due to overlapping of two Dirac cone [8]. The peak position remains same on the Ag island, indicating a very minor charge transfer effect between Ag islands and 3×3 silicene. . Thus, considering the similarity in spectra and the same periodicity, the structure on Ag islands is most likely induced by the interaction between Ag islands and underlying 3×3 silicene film. And this result also proves the preservation of 3×3 silicene beneath Ag islands which further confirms the inert nature 3×3 silicene.

As a comparison, Ag atoms were deposited on √3×√3 silicene film at a substrate temperature of 150 K. Figure 3 shows topography of √3×√3 silicene film after Ag deposition. Similar with Ag deposited on 3×3 silicene film, √3×√3 reconstruction remains same after Ag deposition. However, Ag atoms assemble into nanoclusters, which are neither triangular nor flat top, suggesting a weak interaction between nanoclusters and √3×√3 silicene. This weak interaction is further proved by tip induced movement of Ag islands. STM image in Figure 3(*a*) is taken by scanning from bottom to top. When the STM tip scanned on top of a Ag nanocluster (indicated by a white arrow), the nanocluster suddenly moved toward right direction for around two nanometers. The movement of Ag nanoclusters are also observed in continuously imaging of Figure 3(*c*) and 3(*d*), in which the Ag nanocluster glided and rotated for 45 degree (white arrows in Figure 3(*c*) and 3(*d*) guild to eyes). The movement Similar behavior was reported on FeSe islands grown on chemically inert $sp^2$ hybridized 2D material graphene [19]. Therefore, we believe that √3×√3 silicene may share a similar $sp^2$ dominated hybridization configuration with graphene and may also serve as substrates for high quality sample growth.

To testify the different hybridization states in √7×√7/3×3 silicene and √3×√3 silicene, we applied *in-situ* Raman spectroscopy to investigate phonon modes in the silicene films. Raman spectroscopy is an insightful tool to study the phonon modes associating with the electronic structural properties in 2D materials. The typical Raman spectra of the √7×√7/3×3 silicene film and √3×√3 silicene film are shown in Figure 4. The first-order asymmetric peak located at around 530 cm$^{-1}$ can be interpreted as the zone-centre $E_{2g}$ vibrational mode, which was predicted in previous theoretical studies [20] and demonstrated in experimental results [21, 22]. A shoulder from 495 cm$^{-1}$ to 508 cm$^{-1}$ is ascribed to a quantum confinement effect, which is a common feature also observed in microcrystalline silicon [23] and silicon nanowires [24]. The peak at 230 cm$^{-1}$ is assigned to the "*D*" peak, as its intensity is affected by amount of boundary defects in the √7×√7/3×3 silicene film. There are two major differences between the Raman spectra of √7×√7/3×3 and √3×√3 silicene. Firstly, the "*D*" peak is absent in the Raman results of √3×√3 film, which is consisting with the STM results that less boundaries could be found in the √3×√3 silicene film. Secondly, the much stronger $E_{2g}$ mode was observed in √3×√3 silicene film than that of √7×√7/3×3 silicene film. The $E_{2g}$ vibrational mode corresponds to the bond stretching of all $sp^2$ silicon atoms, and is a fingerprint of honeycomb lattice [21,25,27]. It directly proves that a higher component of $sp^2$ state and a lower component of $sp^3$ state present in √3×√3 silicene than √7×√7/3×3 silicene. It should be noted that electron doping in silicene can result in a hardening of the mode frequency in silicene

[27], *e.g.* $E_{2g}$ mode will shift to higher frequency due to charge doping. As compared to the theoretical simulation [27], a minor electron doping to the epitaxial silicene layers is confirmed.

**4. Summary**

In summary, we systematically investigated the chemical inertness of varies phases of silicene by depositing Ag on silicene films. We found 3×3 silicene is inert to Ag. As a result, no Ag atoms form chemical bonds with it. The more inert nature of √3×√3 silicene is revealed by a weaker interaction between Ag nanoclusters and √3×√3 silicene, leading to Ag nanoclusters gliding on √3×√3 silicene surface. Raman spectroscopy investigations show predominated $sp^2$ hybridization state in √3×√3 silicene which reveals origin of its more inert nature.

**Acknowledgements**

This work is supported by the Australian Research Council (ARC) through a Discovery Project (DP 140102581) and Linkage, Infrastructure, Equipment and Facilities (LIEF) grants (LE100100081 and LE110100099), and partially supported by the National Natural Science Foundation of China (11134005). Topographic images were processed with the WSxM software [26].

**Figure captions**

**Figure 1.** (*a*) STM images of √3×√3 silicene grown on √7×√7/3×3 buffer layer (30 nm × 30 nm, $V_{bias}$ = -1.0 V, $I$ = 1 nA). (*b-d*) High resolution images for 3×3 silicene ($V_{bias}$ = 30 mV, $I$ = 4 nA), √7×√7 silicene ($V_{bias}$ = -20 mV, $I$ = 4 nA), and √3×√3 silicene ($V_{bias}$ = -1 V, $I$ = 3.5 nA) respectively.

**Figure 2.** (*a*) Topography image of 3×3 silicene after deposition of Ag ($V_{bias}$ = 0.8 V, $I$ = 50 pA). Inset figure shows height profile along the solid black line to show the flat top nature of Ag islands. (*b*) High resolution STM image of Ag island in the region denoted by square in (*a*) ($V_{bias}$ = 5 mV, $I$ = 1 nA). (*c*) Tunneling conductance curves along the dotted line in Figure 2(*a*) (from bottom to top; setpoint: $V_{bias}$ = 1.2 V, $I$ = 100 pA)

**Figure 3.** STM topographies of silver growth on silicene √3×√3 surface ($V_{bias}$ = -1 V, $I$ = 100 pA). (*a*) Silver island moved by STM tip during scanning. (*b*, *c*) STM topographies taken in the same area of √3×√3 silicene surface with silver nanoclusters. White lines highlight gliding and rotating of a silver island manipulated by STM tip.

**Figure 4.** Raman spectra of √7×√7/3×3 and √3×√3 silicene films. The intensity of the Raman spectra of √7×√7/3×3 silicene film is multiplied by 10 times to make a clear view of the peaks.

**Figure 1**

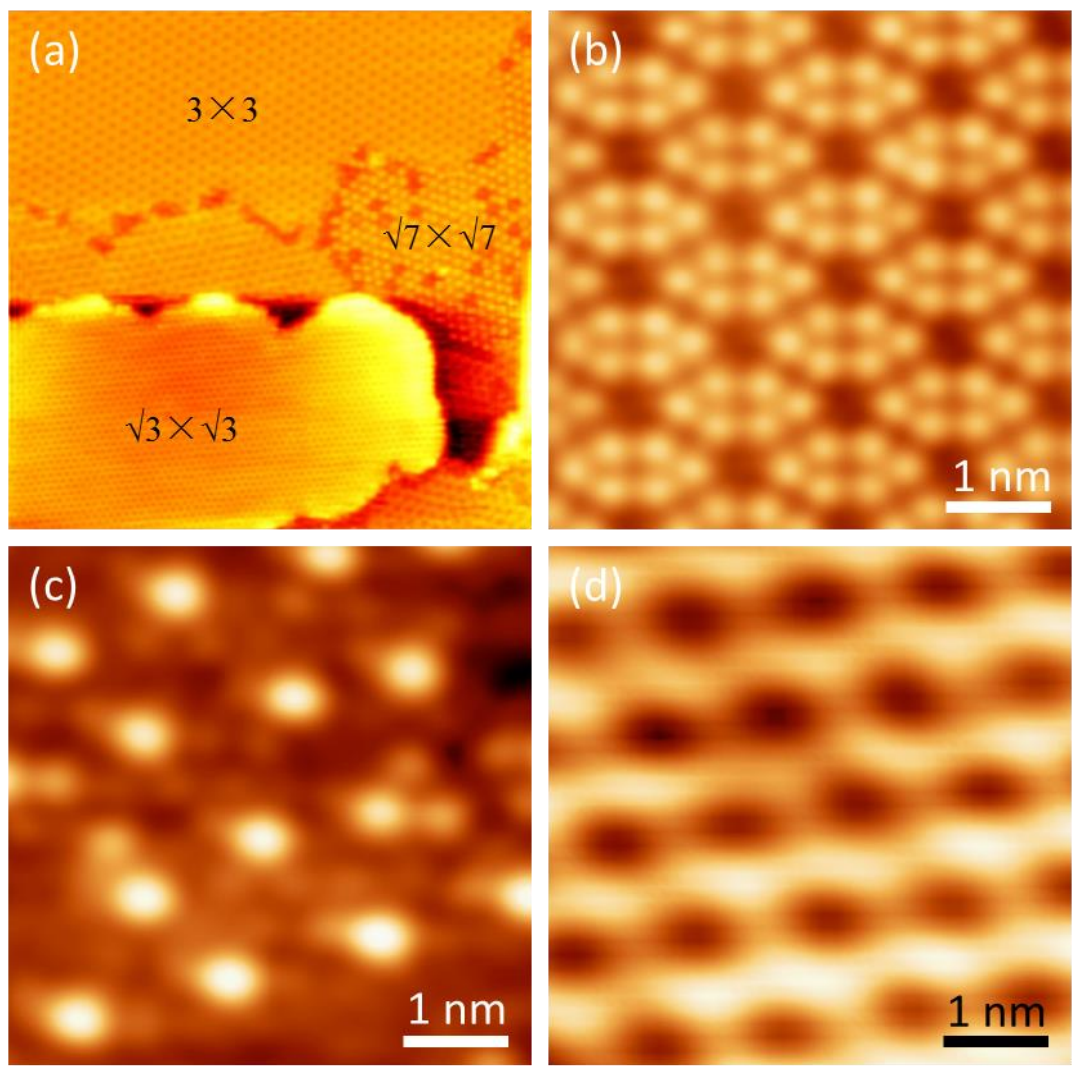

**Figure 2**

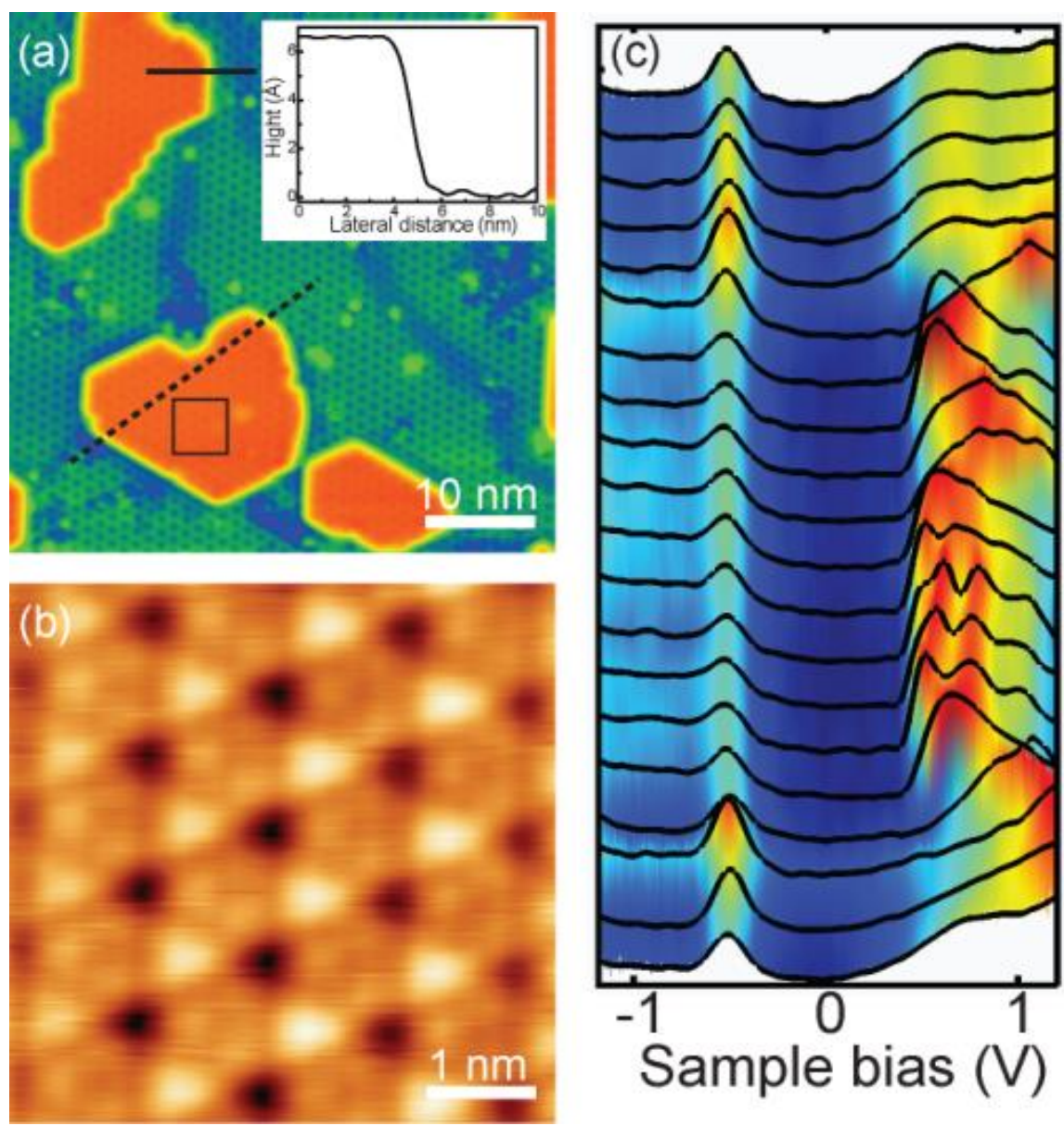

**Figure 3**

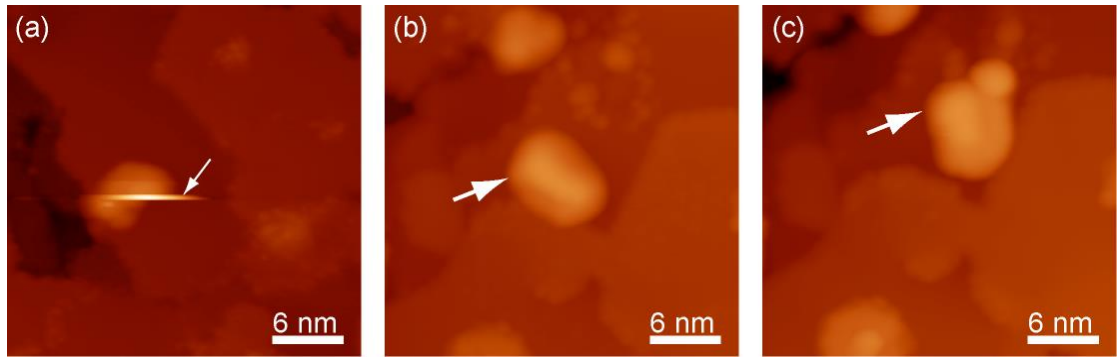

**Figure 4**

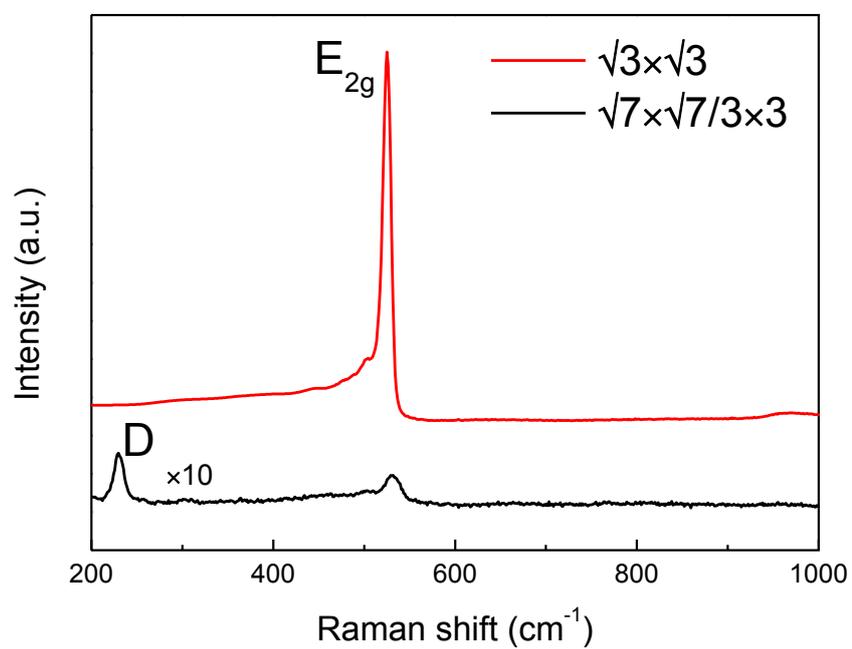